# Extracting random numbers from quantum tunnelling through a single diode

Ramón Bernardo-Gavito[1], Ibrahim Ethem Bagci[2], Jonathan Roberts[1], James Sexton[3], Benjamin Astbury[1], Hamzah Shokeir[1], Thomas McGrath[1], Yasir J. Noori[1], Christopher S. Woodhead[1], Mohamed Missous[2], Utz Roedig[2] and Robert J. Young[1,*]

**Random number generation is crucial in many aspects of everyday life, as online security and privacy depend ultimately on the quality of random numbers. Many current implementations are based on pseudo-random number generators, but information security requires true random numbers for sensitive applications like key generation in banking, defence or even social media. True random number generators are systems whose outputs cannot be determined, even if their internal structure and response history are known. Sources of quantum noise are thus ideal for this application due to their intrinsic uncertainty. In this work, we propose using resonant tunnelling diodes as practical true random number generators based on a quantum mechanical effect. The output of the proposed devices can be directly used as a random stream of bits or can be further distilled using randomness extraction algorithms, depending on the application.**

Random number generators (RNGs) are important in diverse applications such as cryptography, simulations, testing, address generation, and gaming[1]. Many current implementations rely on pseudo-random number generators, but information security requires true random numbers for sensitive applications like key generation in banking, defence or even social media. True random number generators are systems whose outputs cannot be determined, even if their internal structure and response history are known[1]. It has been demonstrated that true random numbers can be obtained from different sources such as noise[2], chaotic systems[3] and quantum phenomena[4]. The main advantage of using sources of quantum noise is its intrinsic uncertainty, as opposed to the predictability of classical sources of noise. In this work, we propose using quantum tunnelling in a simple semiconductor structure, namely a resonant tunnelling diode (RTD). These devices are practical and scalable sources of randomness whose behaviour is governed by quantum physics at room temperature. The semiconductor nature of RTD's, and the simple system proposed to read random numbers from them, makes them a promising candidate for integration into microelectronic systems. The potential to integrate single elements RNGs into current technologies makes them resistant to frequency injection and biasing attacks, which affect state-of-the-art RNGs such as those based on free running oscillators[5]. The output of these devices can be directly used as a random stream of bits or can be further distilled using randomness extraction algorithms, depending on the application.

Resonant tunnelling diodes are the technological realisation of a semiconductor quantum well (QW) with finite rectangular barriers[6,7]. They consist of a thin, narrow band-gap semiconductor structure acting as a quantum well between two wide band-gap semiconductor tunnelling barriers[8,9]. Beyond the tunnelling barriers, highly doped regions of the narrow band-gap semiconductor are usually referred to as the emitter/collector regions, analogous to those in traditional bipolar transistors. Recently, resonant tunnelling devices using quantum dots[10,11], atomic-scale defects[12], graphene[13,14] and other two-dimensional materials[15,16] have been demonstrated, and this has renewed the interest in investigating resonant tunnelling and its applications using new materials. A high-resolution image of a typical RTD used in this work[8], consisting of a square mesa (containing the quantum well structure) and an air bridge (for electrical connection), can be seen in Figure 1-a.

When swept with a DC voltage source, RTDs show the characteristic N-shaped I-V curve of negative differential resistance devices. A typical RTD characteristic is shown in Figure 1-b. The current obtained on the first slope of the curve arises due to the resonant tunnelling process[6,17] that gives this device its name. Here, the bias voltage has shifted the first QW state between the emitter's Fermi level and the lower edge of its conduction band, facilitating the resonant tunnelling of electrons and allowing a reasonably high current to be measured. Once the confined level falls below the conduction band, a sudden drop in current can be observed, as there are no more occupied states in the emitter aligned with the QW state. Further increase of the voltage leads to an increase in the current due to other conduction processes such as thermionic emission of hot electrons over the two tunnelling barriers[17]. The sudden current drop occurring between the two conduction regimes appears as a narrow resonance in the I-V characteristics. The bistability[18] and fast switching characteristics[19] emerging from the sharp NDR resonance of RTDs make them promising candidates for applications in multi-valued logic circuits[20], random-access memories[21], multi-function logic gates[22], chaotic signal generation[3,23,24], single-photon switching[25], unique device identification[26] and terahertz oscillators[27,28].

[1] Physics Department, Lancaster University, Lancaster, LA1 4YB, UK
[2] School of Computing and Communications, Lancaster University, Lancaster, LA1 4WA, UK.
[3] School of Electrical and Electronic Engineering, University of Manchester, M13 9PL, UK.
[*] Correspondence should be addressed to R.J.Y. (email: r.j.young@lancaster.ac.uk)

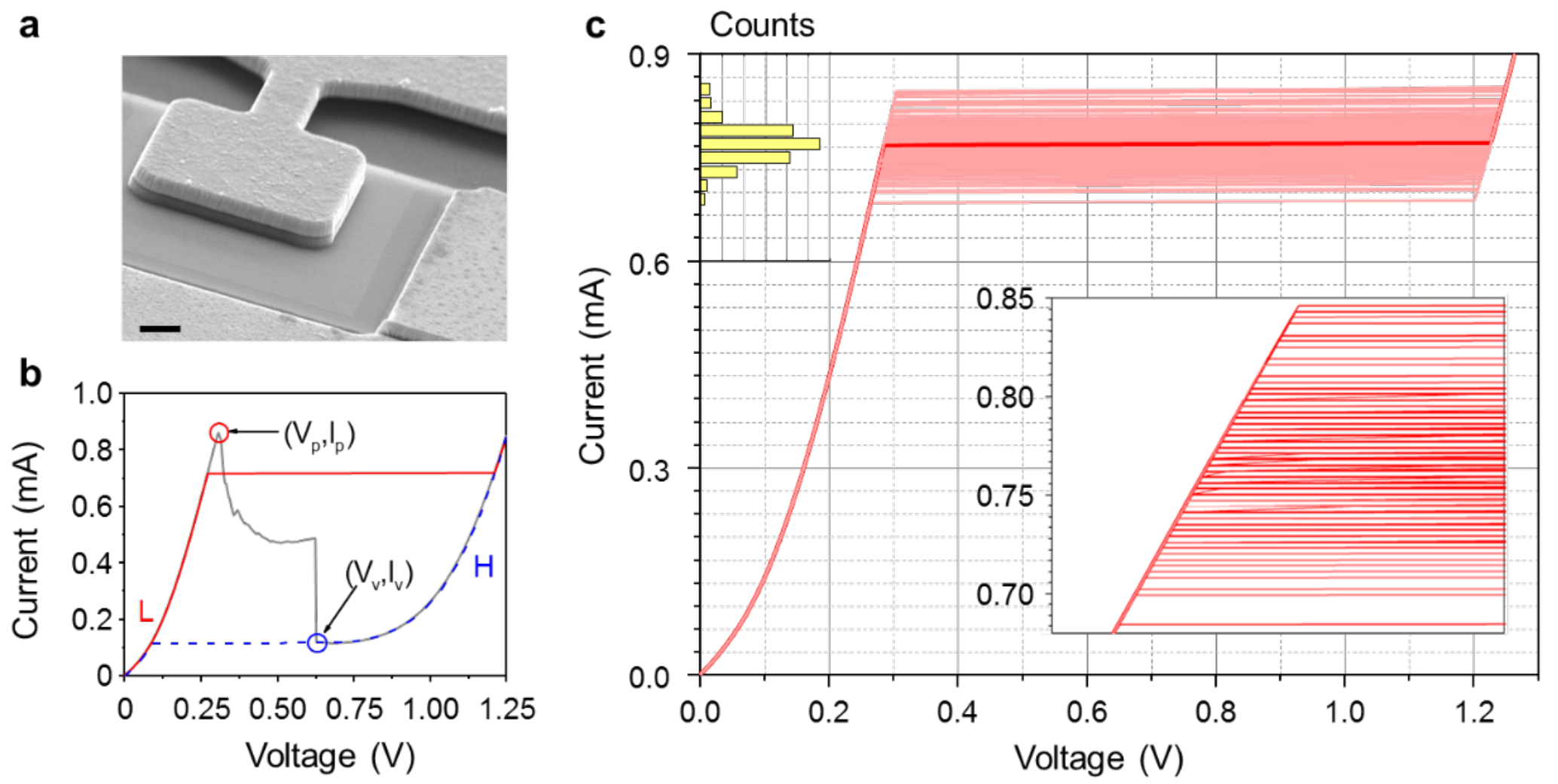

**Figure 1. Physical and electrical characteristics of RTDs. (a)** SEM micrograph of one of the studied 25 μm$^2$ RTDs taken with a tilt of 20 degrees (scale bar 1 μm). **(b)** I-V characteristic of a 4 μm$^2$ RTD showing the NDR curve with a voltage sweep (grey line) and the hysteretic behaviour with current sweeps. A jump from the low-resistance state to the high-resistance state can be observed in forward sweeps (red line) near the peak, and the opposite will happen in reverse sweeps (blue line) near the valley. **(c)** 100 forward current sweeps (light red lines) showing the random distribution of state changes around an average value (bold red line). This average curve corresponds to the red line shown in (b). The histogram shows the switching probability as a function of current (same vertical units as the main axis). The inset shows a zoom-in to the switching region.

Due to their N-shaped NDR characteristics, the RTD's current is a single-valued function of voltage whereas the opposite is not true, as current values between those of the peak ($I_p$) and the valley ($I_v$) exhibit multiple voltage levels[18] due to the different conduction mechanisms. Using a current source in that range can result in two different scenarios: a low-resistance state (L) corresponding to the first positive differential resistance (PDR) region, and a high-resistance state (H) corresponding to the second PDR region. The instability of the NDR region[29,30] prevents the system from staying in that voltage range for long times, pushing the system to one of the two PDR branches. Ramping the current up and down results in a hysteresis cycle[31,32] between these two resistance states as shown in Figure 1-b.

RESULTS AND DISCUSSION

Figure 1 shows that sweeping current across the device results in a hysteresis cycle. The forward sweep takes place in the first PDR region up to a threshold near the peak current where the voltage is pushed to the second PDR region of the curve. Once that threshold is passed, ramping down the current will not push the voltage back to the first PDR region until the valley current is reached. Our measurements show that the switching threshold from one state to the other is not a fixed value but is statistically distributed near the resonance current, $I_p$ (Figure 1-c). As expected, currents above $I_p$ will always send the system to the second PDR region as the low-resistance branch does not reach that current range, and setting the current below $I_v$ will always give the low resistance state for similar reasons. On the other hand, working between $I_v$ and $I_p$ results in a non-deterministic switching behaviour, as the state change from one slope to the other happens at different values each time following a probability distribution as shown in the histogram of Figure 1-c. The origin of the uncertainty on the response of the RTD to current sources is most likely related to the charge build-up and trap filling in the quantum well and adjoining regions. This leads to shifts in the energy of the confined level[6,18], dynamically altering the threshold at which resonant tunnelling occurs.

To further characterise this stochastic switching behaviour and exploit its possibilities as an eventual random number generator, we performed a series of experiments using current pulse trains of varying amplitude and frequency. A Keithley 2602B source-measure unit (SMU) is programmed to send periodic current pulses of a fixed amplitude and length. For the sake of simplicity and keeping a small parameter space, the duty cycle of the pulse trains is kept at 50%, while varying the amplitude and pulse width. The voltage drop across the RTD is measured with the SMU at the end of each pulse, resulting in a value either in the first or second PDR regions. Alternatively, the output voltage can be measured using a fast oscilloscope to characterise the time response of the system.

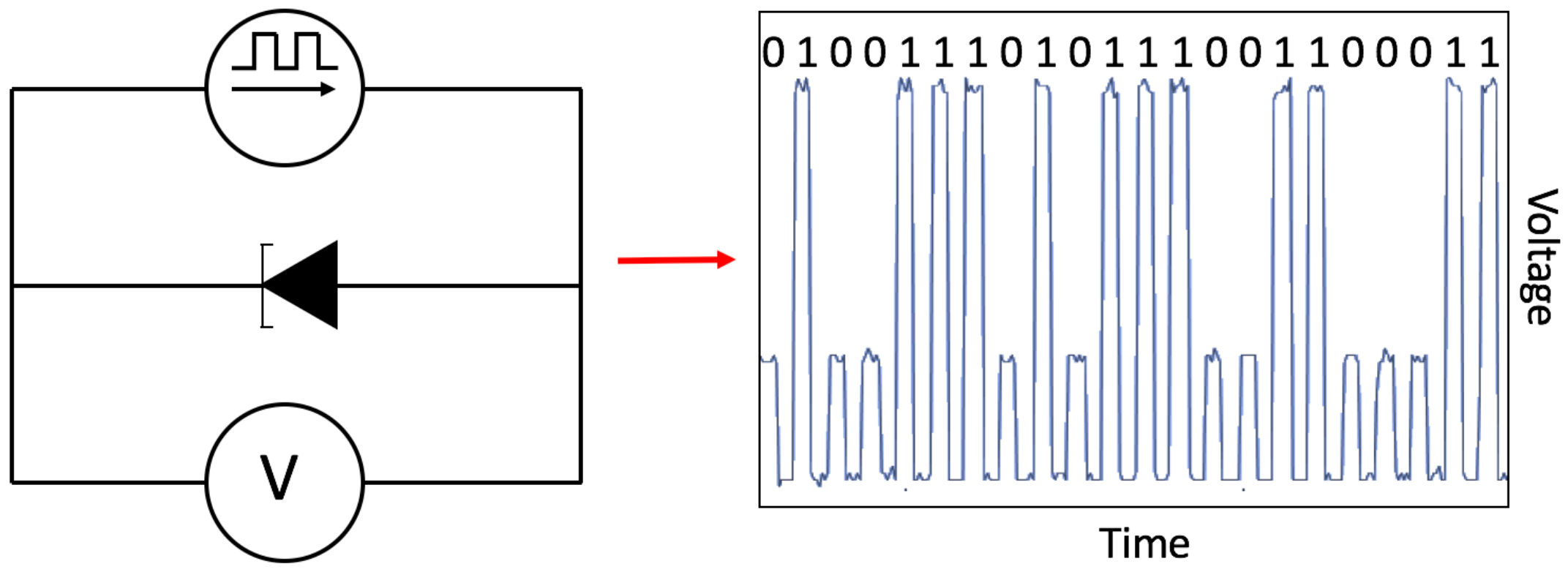

**Figure 2. The experimental arrangement used to generate random numbers.** A pulsed current source drives the RTD while the voltage across it is measured. The right panel shows an oscilloscope trace illustrating the random response, along with the corresponding logical levels.

Analysing the time response of different RTDs to current pulse trains shows that the random behaviour of the switching threshold is a dynamic process. This relates to the time that it takes the system to jump between the first and second PDR regions. When the source is set at a fixed current, the device will start conducting in the first PDR region of the I-V curve, i.e. that dominated by resonant tunnelling through the first quantum well state, and after a period of time it will jump into the second PDR region. This dynamic switching behaviour can be related to the charge accumulation in the quantum well or in charge traps distributed along the structure, which are known to have an influence in the behaviour of RTDs[6,18]. The charge and discharge of these features lead to a shift in the energy of the confined quantum level, thus allowing switching from one tunnelling mechanism to another. The charge accumulation in the quantum well and charge traps is known to be a dynamic process that can take from picoseconds to hundreds of milliseconds[6], which fits well with our experimental results.

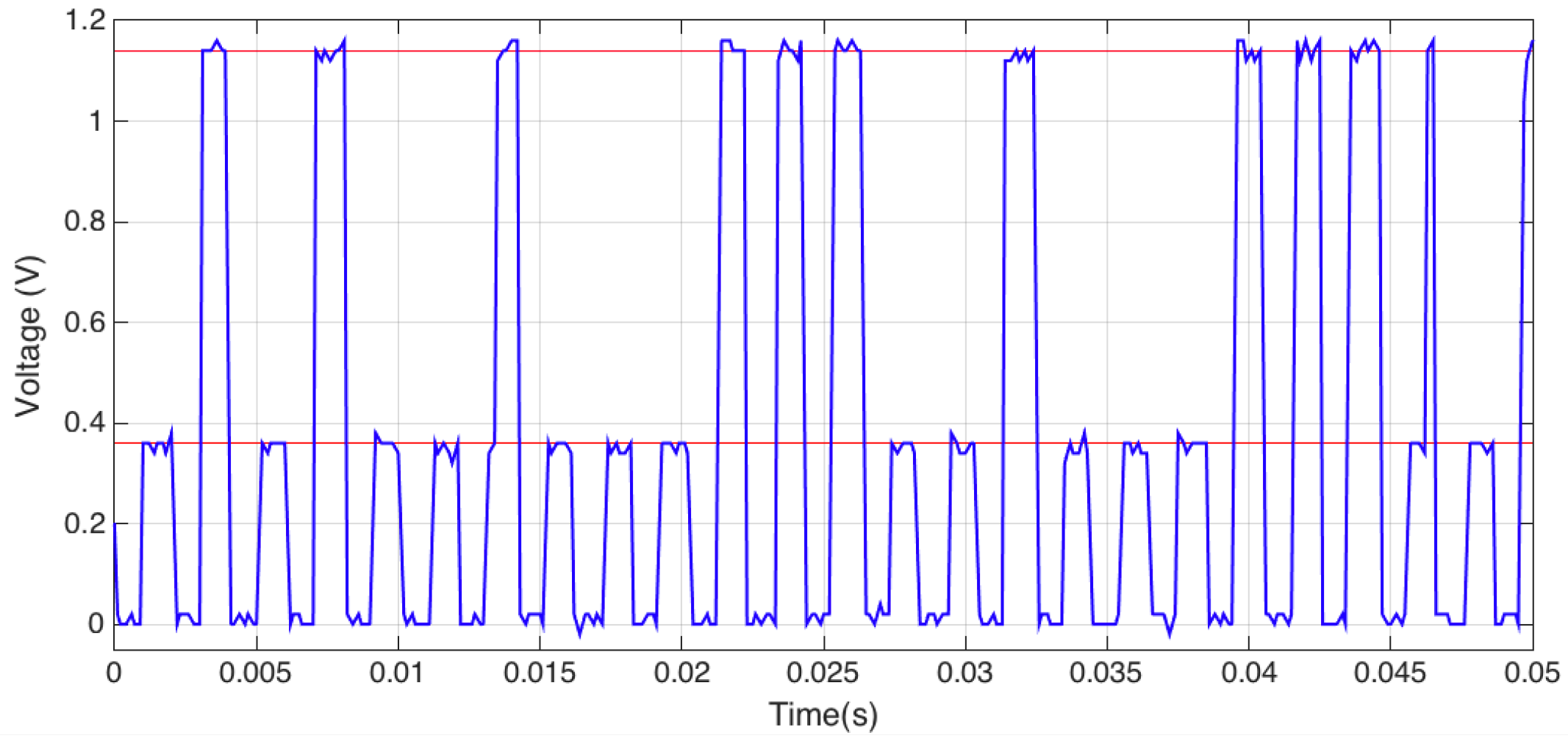

**Figure 3. Time dependent voltage measurements of an RTD driven with a pulse train of 1.50 mA, 1 ms pulse width, 50% duty cycle.** The red lines mark the position of the LOW and HIGH levels. The voltage pulses measured at 13ms and 46ms show the transition from LOW to HIGH during the corresponding current pulse.

In order to operate RTDs as sources of randomness, we can exploit the aforementioned random dynamic switching to obtain a stream of random bits. By sourcing current pulses to an RTD and measuring the voltage drop across the device at a given time, i.e. at the end of each pulse, we will obtain a value that corresponds either to the first or second PDR regions. To simplify the description of the operation we will denote these two states as L and H respectively. The experiment shows that the L/H ratio depends both on the current level and the pulse width, which allows us to easily change the probability distribution of the output.

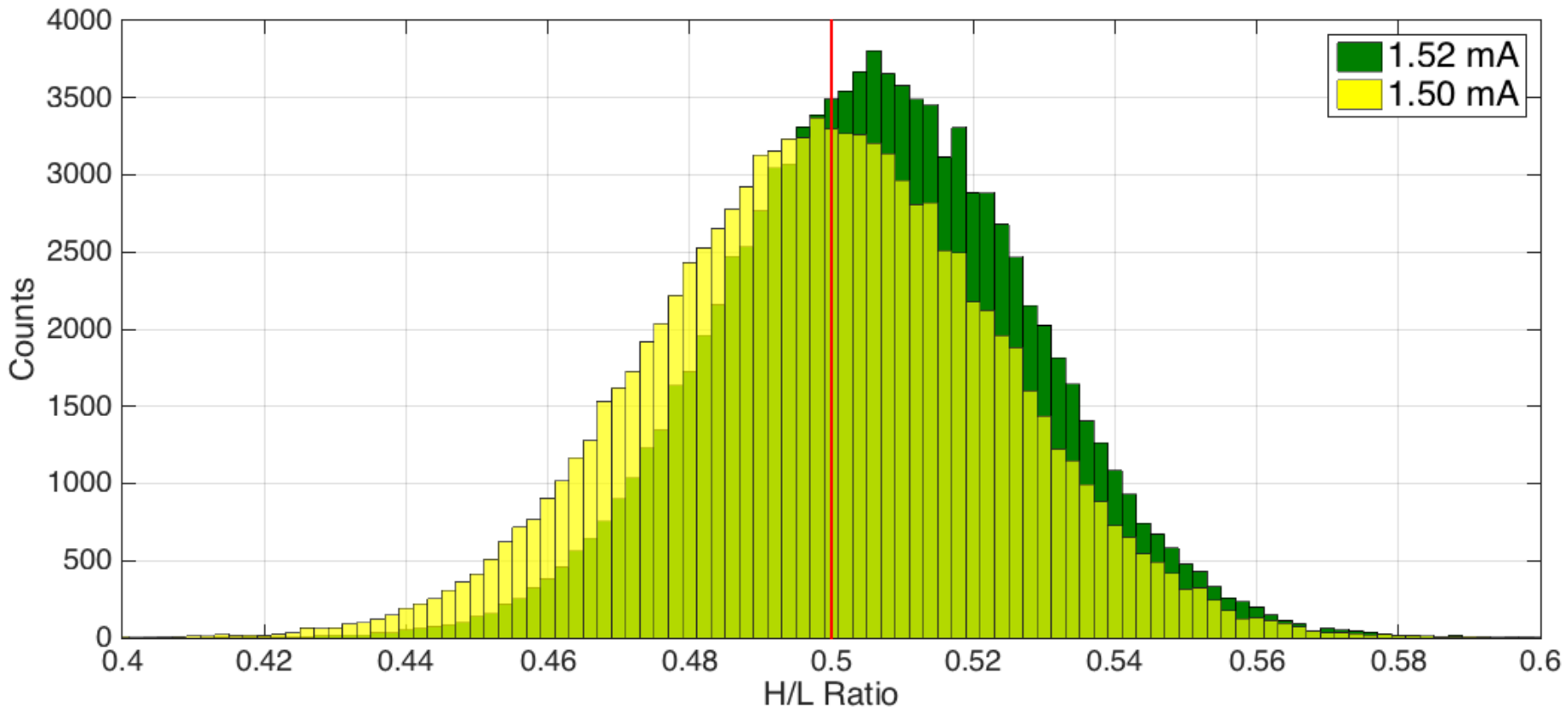

**Figure 4. High to low ratio tuning as a function of pulse amplitude.** The high to low ratio of the RTD can be tuned by changing the current amplitude of the pulses. The graph shows two histograms of the output distribution of two different pulse trains, 1ms wide, 50% duty cycle with different current amplitudes, namely 1.50 mA (yellow) and 1.53 mA (dark green). The red line marks the 50/50 point. The results show how we can tune the average output of a RTD-based RNG by changing the current level. Each histogram corresponds to a sample size of $5x10^7$ pulses divided in subsets of 500 pulses.

Considering the dynamic behaviour described above, the output will be a random distribution of L and H values with characteristics depending upon the amplitude and width of the pulses. For low currents and short pulses the distribution will be strongly biased towards L, while high currents and long times will make the output more likely to be H. Tuning these two parameters allows us to set the probability distribution of the output. This can be explained by considering that the charge trapping, which is responsible for the switching between the two conduction mechanisms, is a probabilistic effect itself. A higher current corresponds to a larger number of electrons crossing the device in a given time, so more charge can be trapped, assuming a fixed probability. Likewise, if the pulse is longer, it is more probable that at the end of the pulse the trapped charge is enough to push the confined level to higher energies and facilitate the alternative conduction path.

Most applications employing random numbers, such as cryptography, require an unbiased uniform output distribution. In other cases, shifted or skewed distributions might be required. For example, a shifted probability distribution can be useful for simulating the random path of a particle subject to a certain potential, or they may be used in gambling or stock market predictions. One potential advantage of an RTD-based random number generator is that its average output can be modified *in operando*, which can be exploited for more complex simulations.

Our experiments show that, although the underlying principle of operation of the RTD as an RNG is the quantum tunnelling through the resonant structure, it is convoluted with classical thermal noise and other types of noises from the measurement equipment. Variations in the environment temperature and electrical noise can influence the output of the proposed RNG. Our experiments show a slow drift correlated to the evolution of the room temperature. Although this effect can be considered to add up to the randomness itself, as it will push the L/H ratio out of the set point in a random way, it is undesirable for most applications. Also, it adds a strong source of classical noise to a device that is intended to be used as a quantum random number generator. The use of a simple feedback mechanism to correct for temperature swings mitigates the effect of thermal drift, as it is a very slow process compared to pulse period. Regardless, the operating principle discussed still involves a classical measurement of a quantum process, making the system subject to any classical sources of noise that can obfuscate the pure quantum randomness.

The problem of discriminating between classical and quantum randomness sources has already been addressed in previous work, and there exist randomness extraction algorithms specifically designed to distil the output from quantum RNGs to obtain uniformly distributed random numbers. As an example, we used the double-hash function algorithm suggested by Frauchiger et al.[33], which is computationally efficient and could be implemented in hardware. The resulting data successfully passes the 15 tests in the NIST randomness test suite[34] with a significance level of 0.05 (5%). Specific details on testing are given in the supplementary information.

CONCLUSION

In this paper, we have shown how the uncertainty in the switching between two conduction mechanisms in a resonant tunnelling diode can be exploited to produce a random number generator. The operating voltages of the two conduction states can be easily mapped to digital logic to generate a random bit stream. Although the raw output of the proposed scheme is still affected by classical noise, it can be used for many applications requiring random numbers. Distilling the raw output using a double-hash function to correct the effect of the classic environment results in bit streams that complies with the NIST suite of randomness tests, a standard in random number generator testing for cryptographic applications.

One of the advantages of RTDs used in this study is that the operating voltage levels can be easily interfaced with logic levels in microelectronics. In the example above, the first PDR region lays below 0.4V, i.e. before the resonance. At the working currents this means that the L level will be close to this voltage. The corresponding H level for that current projected to the second PDR region corresponds to around 1.15V.

ACKNOWLEDGEMENTS

RJY acknowledges support by the Royal Society through a University Research Fellowship (UF110555 and UF160721). This material is based upon work supported by the Air Force Office of Scientific Research under award number FA9550-16-1-0276. This work was also supported by grants from The Engineering and Physical Sciences Research Council in the UK (EP/K50421X/1 and EP/L01548X/1), and the Royal Society through a Brian Mercer award.

AUTHOR CONTRIBUTIONS

RJY, UR and RB-G designed the project. MM and JS fabricated and performed initial tests on the RTD devices. RB-G, JR, CSW, BA and HS performed the detailed electronic measurements. RB-G, IEB, TM and YJN analysed the data. The manuscript was prepared primarily by RB-G, and RJY with contribution from all authors.

**Supplementary information: Extracting random numbers from quantum tunnelling through a single diode**

Randomness extraction

Frauchiger et al.[1] provide a framework to remove negative effects of imperfect quantum processes on QRNGs. Practical implementations of QRNGs are always subject to noise. This noise cannot be controlled and is not guaranteed to be random. Proposed framework applies randomness extraction to the result of the device to get true randomness. Randomness extraction consists of block-wise two-universal hashing. Here the result of the device is divided into n-bit sized blocks, and each block is given to a hash function to get l-bit sized new blocks. New blocks are then concatenated to get the final result, and l is generally smaller than n. With the careful selection of the parameters n and l as well as two-universal hashing algorithm, truly random blocks can be generated. The paper includes efficient implementation of proposed randomness extraction, which we used in this work.

NIST tests

In order to check the viability of the proposed random number generators as a part of a cryptographic scheme we performed the battery of tests included in the NIST randomness test suite[2]. Running the tests on the raw data extracted directly from the RTD shows that the sequences of bits are locally random but present a long-term lack of randomness due to the thermal fluctuations described in the main text.

On the other hand, when the data is distilled using the double hash function method we obtain results that are compliant with the NIST randomness standards for cryptographic applications. The code used for the randomness extraction is described in the previous section. Table S1 (provided as a MS Excel spreadsheet) shows the results of the tests ran on $50\times10^6$ raw bits. The output after applying the randomness extraction code consists in $16.5\times10^6$ bits and is divided in 30 datasets of 550000 bits.

For the longest runs of ones test we used N=100 substrings with a length of M=10000 bits. Non-overlapping templates test was ran using a sequence length of n=1000000, block length m=9, substring length M=125000, N=8 substrings. For the overlapping template of all ones tests, these values were n=1000000, m=9, M=1032, and N=968. The linear complexity test used a substring length of M=500 and N=2000 substrings. The sequence length for the random excursion test and its variant was n=1000000, with varying number of cycles depending on the sequence. The rank tests use its default value of 97 matrices for the given data size. Serial test used a block length of m=16 for sequences of n=1000000 bits. The Mauer's universal statistical test used the default values of L=7, Q=1280, and K=141577 for the given input. For the universal entropy test we used a block length of m=10 bits with sequences of n=1000000 bits. The significance level was set to $\alpha=0.05$ (5%), as it is the most widely used value for this kind of tests.

As stated in the NIST randomness tests suite, he minimum pass rate for each statistical test except for the random excursion (variant) test is approximately is 24 for a sample size of 30 binary sequences. The minimum pass rate for the random excursion (variant) test
is approximately 10 for a sample size of 14 binary sequences. Al the rests that were run are successful considering both the proportion and the P-value criteria as described in the NIST randomness test suite manual[2].

RESULTS FOR THE UNIFORMITY OF P-VALUES AND THE PROPORTION OF PASSING SEQUENCES

| C1 | C2 | C3 | C4 | C5 | C6 | C7 | C8 | C9 | C10 | P-VALUE | PROPORTION | STATISTICAL TEST |
|---|---|---|---|---|---|---|---|---|---|---|---|---|
| 2 | 1 | 6 | 4 | 3 | 2 | 3 | 3 | 6 | 0 | 0.253551 | 28/30 | Frequency |
| 3 | 4 | 2 | 1 | 5 | 2 | 1 | 5 | 6 | 1 | 0.299251 | 29/30 | BlockFrequency |
| 5 | 1 | 2 | 3 | 2 | 2 | 6 | 4 | 4 | 1 | 0.468595 | 28/30 | CumulativeSums |
| 2 | 2 | 2 | 2 | 4 | 6 | 1 | 2 | 6 | 3 | 0.407091 | 29/30 | CumulativeSums |
| 0 | 2 | 2 | 2 | 2 | 0 | 5 | 1 | 6 | 10 | 0.000569 | 30/30 | Runs |
| 3 | 5 | 2 | 3 | 3 | 0 | 5 | 3 | 3 | 3 | 0.739918 | 30/30 | LongestRun |
| 2 | 3 | 2 | 3 | 5 | 0 | 6 | 3 | 3 | 3 | 0.534146 | 28/30 | Rank |
| 5 | 3 | 4 | 2 | 4 | 2 | 1 | 2 | 2 | 5 | 0.739918 | 28/30 | FFT |
| 7 | 2 | 2 | 2 | 4 | 2 | 3 | 1 | 4 | 3 | 0.468595 | 27/30 | NonOverlappingTemplate |
| 4 | 3 | 4 | 1 | 1 | 3 | 6 | 0 | 5 | 3 | 0.299251 | 29/30 | NonOverlappingTemplate |
| 2 | 3 | 0 | 1 | 2 | 8 | 5 | 4 | 3 | 2 | 0.082177 | 30/30 | NonOverlappingTemplate |
| 5 | 0 | 2 | 3 | 2 | 3 | 0 | 4 | 5 | 6 | 0.178278 | 27/30 | NonOverlappingTemplate |
| 7 | 3 | 2 | 0 | 3 | 2 | 2 | 1 | 5 | 5 | 0.148094 | 28/30 | NonOverlappingTemplate |
| 2 | 1 | 2 | 4 | 3 | 2 | 4 | 9 | 3 | 0 | 0.035174 | 29/30 | NonOverlappingTemplate |
| 2 | 6 | 4 | 3 | 5 | 4 | 1 | 1 | 1 | 3 | 0.407091 | 30/30 | NonOverlappingTemplate |
| 1 | 4 | 4 | 4 | 2 | 4 | 1 | 1 | 7 | 2 | 0.253551 | 29/30 | NonOverlappingTemplate |
| 5 | 2 | 1 | 1 | 5 | 3 | 4 | 6 | 3 | 0 | 0.213309 | 28/30 | NonOverlappingTemplate |
| 1 | 2 | 5 | 1 | 4 | 1 | 2 | 5 | 3 | 6 | 0.299251 | 29/30 | NonOverlappingTemplate |
| 2 | 6 | 1 | 0 | 4 | 7 | 2 | 1 | 3 | 4 | 0.082177 | 29/30 | NonOverlappingTemplate |
| 2 | 2 | 3 | 3 | 3 | 3 | 5 | 5 | 4 | 0 | 0.671779 | 30/30 | NonOverlappingTemplate |
| 1 | 0 | 2 | 3 | 6 | 5 | 0 | 5 | 3 | 5 | 0.100508 | 30/30 | NonOverlappingTemplate |
| 5 | 2 | 3 | 4 | 1 | 4 | 2 | 3 | 3 | 3 | 0.911413 | 27/30 | NonOverlappingTemplate |
| 2 | 4 | 3 | 2 | 3 | 3 | 3 | 1 | 1 | 8 | 0.213309 | 29/30 | NonOverlappingTemplate |
| 3 | 6 | 3 | 1 | 3 | 3 | 3 | 5 | 1 | 2 | 0.602458 | 27/30 | NonOverlappingTemplate |
| 3 | 6 | 4 | 3 | 2 | 2 | 1 | 2 | 5 | 2 | 0.602458 | 28/30 | NonOverlappingTemplate |
| 3 | 2 | 3 | 1 | 6 | 2 | 4 | 4 | 2 | 3 | 0.739918 | 27/30 | NonOverlappingTemplate |
| 2 | 1 | 2 | 3 | 3 | 4 | 2 | 3 | 4 | 6 | 0.739918 | 28/30 | NonOverlappingTemplate |
| 2 | 4 | 3 | 6 | 2 | 7 | 0 | 3 | 1 | 2 | 0.122325 | 30/30 | NonOverlappingTemplate |
| 3 | 0 | 3 | 5 | 3 | 2 | 3 | 6 | 3 | 2 | 0.534146 | 29/30 | NonOverlappingTemplate |
| 4 | 4 | 1 | 0 | 2 | 3 | 4 | 4 | 5 | 3 | 0.602458 | 28/30 | NonOverlappingTemplate |
| 1 | 5 | 1 | 2 | 2 | 6 | 3 | 3 | 1 | 6 | 0.213309 | 30/30 | NonOverlappingTemplate |
| 2 | 1 | 6 | 4 | 5 | 2 | 4 | 0 | 1 | 5 | 0.178278 | 29/30 | NonOverlappingTemplate |
| 4 | 2 | 2 | 2 | 5 | 4 | 4 | 2 | 3 | 2 | 0.911413 | 27/30 | NonOverlappingTemplate |
| 3 | 2 | 2 | 2 | 4 | 3 | 3 | 3 | 4 | 4 | 0.991468 | 27/30 | NonOverlappingTemplate |
| 4 | 2 | 1 | 6 | 2 | 0 | 1 | 7 | 6 | 1 | 0.022503 | 27/30 | NonOverlappingTemplate |
| 1 | 4 | 1 | 2 | 3 | 3 | 2 | 7 | 3 | 4 | 0.407091 | 29/30 | NonOverlappingTemplate |
| 1 | 3 | 4 | 0 | 4 | 4 | 5 | 4 | 2 | 3 | 0.602458 | 29/30 | NonOverlappingTemplate |
| 5 | 2 | 1 | 3 | 3 | 2 | 7 | 4 | 1 | 2 | 0.299251 | 26/30 | NonOverlappingTemplate |
| 3 | 3 | 2 | 6 | 0 | 3 | 4 | 2 | 3 | 4 | 0.602458 | 27/30 | NonOverlappingTemplate |
| 2 | 5 | 4 | 2 | 3 | 2 | 3 | 2 | 4 | 3 | 0.949602 | 29/30 | NonOverlappingTemplate |
| 3 | 4 | 2 | 2 | 5 | 4 | 6 | 1 | 1 | 2 | 0.468595 | 30/30 | NonOverlappingTemplate |
| 2 | 4 | 5 | 1 | 3 | 2 | 4 | 3 | 5 | 1 | 0.671779 | 29/30 | NonOverlappingTemplate |
| 2 | 5 | 2 | 2 | 0 | 4 | 4 | 3 | 2 | 6 | 0.407091 | 29/30 | NonOverlappingTemplate |
| 6 | 1 | 3 | 2 | 0 | 2 | 6 | 5 | 3 | 2 | 0.178278 | 27/30 | NonOverlappingTemplate |
| 4 | 3 | 4 | 4 | 3 | 3 | 2 | 1 | 3 | 3 | 0.97606 | 28/30 | NonOverlappingTemplate |
| 6 | 2 | 2 | 5 | 2 | 4 | 2 | 2 | 2 | 3 | 0.671779 | 26/30 | NonOverlappingTemplate |
| 2 | 3 | 2 | 0 | 4 | 4 | 5 | 4 | 4 | 2 | 0.671779 | 29/30 | NonOverlappingTemplate |
| 1 | 3 | 2 | 4 | 1 | 6 | 2 | 4 | 7 | 0 | 0.082177 | 30/30 | NonOverlappingTemplate |
| 3 | 3 | 3 | 3 | 5 | 1 | 3 | 1 | 6 | 2 | 0.602458 | 28/30 | NonOverlappingTemplate |
| 1 | 3 | 3 | 3 | 1 | 2 | 4 | 4 | 5 | 4 | 0.804337 | 30/30 | NonOverlappingTemplate |
| 3 | 3 | 3 | 1 | 5 | 3 | 5 | 1 | 2 | 4 | 0.739918 | 28/30 | NonOverlappingTemplate |
| 2 | 2 | 5 | 2 | 3 | 3 | 6 | 1 | 2 | 4 | 0.602458 | 29/30 | NonOverlappingTemplate |
| 5 | 6 | 0 | 3 | 5 | 3 | 2 | 2 | 3 | 1 | 0.299251 | 26/30 | NonOverlappingTemplate |
| 3 | 3 | 3 | 5 | 3 | 2 | 2 | 3 | 2 | 4 | 0.97606 | 29/30 | NonOverlappingTemplate |
| 3 | 4 | 4 | 3 | 2 | 3 | 2 | 2 | 2 | 5 | 0.949602 | 30/30 | NonOverlappingTemplate |
| 1 | 5 | 3 | 4 | 0 | 3 | 3 | 1 | 6 | 4 | 0.299251 | 30/30 | NonOverlappingTemplate |
| 3 | 1 | 3 | 3 | 3 | 6 | 1 | 4 | 4 | 2 | 0.671779 | 29/30 | NonOverlappingTemplate |
| 4 | 3 | 2 | 3 | 2 | 1 | 5 | 5 | 3 | 2 | 0.804337 | 28/30 | NonOverlappingTemplate |
| 5 | 2 | 5 | 2 | 2 | 2 | 3 | 3 | 3 | 3 | 0.911413 | 28/30 | NonOverlappingTemplate |
| 5 | 5 | 3 | 1 | 2 | 5 | 2 | 1 | 4 | 2 | 0.534146 | 26/30 | NonOverlappingTemplate |
| 2 | 4 | 2 | 3 | 5 | 4 | 5 | 2 | 1 | 2 | 0.739918 | 30/30 | NonOverlappingTemplate |
| 3 | 0 | 6 | 2 | 3 | 5 | 2 | 4 | 3 | 2 | 0.468595 | 28/30 | NonOverlappingTemplate |
| 4 | 2 | 1 | 3 | 3 | 2 | 6 | 2 | 1 | 6 | 0.350485 | 29/30 | NonOverlappingTemplate |
| 4 | 5 | 5 | 6 | 2 | 3 | 1 | 0 | 2 | 2 | 0.253551 | 29/30 | NonOverlappingTemplate |
| 4 | 2 | 1 | 2 | 4 | 2 | 3 | 4 | 5 | 3 | 0.862344 | 29/30 | NonOverlappingTemplate |
| 5 | 6 | 1 | 2 | 4 | 4 | 2 | 2 | 3 | 1 | 0.468595 | 26/30 | NonOverlappingTemplate |
| 1 | 5 | 2 | 6 | 4 | 0 | 3 | 1 | 2 | 6 | 0.122325 | 30/30 | NonOverlappingTemplate |
| 5 | 0 | 2 | 3 | 4 | 3 | 5 | 3 | 2 | 3 | 0.671779 | 28/30 | NonOverlappingTemplate |
| 6 | 2 | 4 | 5 | 6 | 1 | 4 | 0 | 1 | 1 | 0.082177 | 25/30 | NonOverlappingTemplate |
| 2 | 2 | 2 | 9 | 2 | 1 | 3 | 5 | 0 | 4 | 0.022503 | 28/30 | NonOverlappingTemplate |
| 2 | 4 | 1 | 2 | 2 | 3 | 5 | 4 | 3 | 4 | 0.862344 | 28/30 | NonOverlappingTemplate |
| 0 | 6 | 3 | 0 | 6 | 4 | 2 | 1 | 4 | 4 | 0.100508 | 30/30 | NonOverlappingTemplate |
| 3 | 4 | 3 | 1 | 5 | 3 | 2 | 1 | 2 | 6 | 0.534146 | 29/30 | NonOverlappingTemplate |
| 2 | 1 | 2 | 1 | 4 | 2 | 4 | 4 | 5 | 5 | 0.602458 | 29/30 | NonOverlappingTemplate |
| 2 | 2 | 3 | 4 | 4 | 4 | 1 | 5 | 1 | 4 | 0.739918 | 29/30 | NonOverlappingTemplate |
| 4 | 3 | 4 | 3 | 1 | 4 | 0 | 4 | 3 | 4 | 0.739918 | 28/30 | NonOverlappingTemplate |
| 2 | 1 | 3 | 5 | 4 | 1 | 2 | 3 | 5 | 4 | 0.671779 | 29/30 | NonOverlappingTemplate |
| 4 | 4 | 1 | 1 | 2 | 7 | 2 | 2 | 0 | 7 | 0.035174 | 29/30 | NonOverlappingTemplate |
| 2 | 6 | 2 | 3 | 5 | 3 | 1 | 2 | 1 | 5 | 0.407091 | 30/30 | NonOverlappingTemplate |
| 2 | 5 | 4 | 4 | 2 | 3 | 4 | 0 | 4 | 2 | 0.671779 | 30/30 | NonOverlappingTemplate |
| 6 | 6 | 1 | 2 | 2 | 2 | 3 | 3 | 3 | 2 | 0.468595 | 27/30 | NonOverlappingTemplate |
| 3 | 1 | 1 | 4 | 0 | 7 | 2 | 5 | 3 | 4 | 0.148094 | 30/30 | NonOverlappingTemplate |
| 7 | 2 | 2 | 2 | 4 | 2 | 3 | 1 | 4 | 3 | 0.468595 | 27/30 | NonOverlappingTemplate |
| 1 | 4 | 4 | 3 | 2 | 0 | 3 | 2 | 5 | 6 | 0.350485 | 29/30 | NonOverlappingTemplate |
| 2 | 2 | 4 | 4 | 5 | 1 | 3 | 2 | 5 | 2 | 0.739918 | 29/30 | NonOverlappingTemplate |
| 2 | 4 | 2 | 2 | 3 | 4 | 2 | 2 | 3 | 6 | 0.804337 | 30/30 | NonOverlappingTemplate |
| 4 | 5 | 1 | 3 | 1 | 3 | 5 | 2 | 3 | 3 | 0.739918 | 29/30 | NonOverlappingTemplate |
| 3 | 2 | 6 | 4 | 1 | 4 | 2 | 1 | 4 | 3 | 0.602458 | 28/30 | NonOverlappingTemplate |
| 6 | 2 | 3 | 1 | 2 | 0 | 6 | 3 | 3 | 4 | 0.253551 | 27/30 | NonOverlappingTemplate |
| 1 | 5 | 3 | 2 | 2 | 5 | 5 | 3 | 1 | 3 | 0.602458 | 30/30 | NonOverlappingTemplate |
| 2 | 4 | 5 | 5 | 1 | 6 | 2 | 3 | 1 | 1 | 0.299251 | 29/30 | NonOverlappingTemplate |
| 2 | 5 | 1 | 3 | 4 | 1 | 3 | 3 | 6 | 2 | 0.534146 | 29/30 | NonOverlappingTemplate |
| 1 | 3 | 4 | 5 | 3 | 3 | 4 | 3 | 2 | 2 | 0.911413 | 30/30 | NonOverlappingTemplate |
| 2 | 4 | 4 | 2 | 1 | 1 | 3 | 3 | 4 | 6 | 0.602458 | 30/30 | NonOverlappingTemplate |
| 3 | 2 | 5 | 3 | 1 | 5 | 3 | 3 | 3 | 2 | 0.862344 | 28/30 | NonOverlappingTemplate |
| 2 | 3 | 3 | 4 | 5 | 2 | 2 | 2 | 4 | 3 | 0.949602 | 28/30 | NonOverlappingTemplate |
| 4 | 2 | 1 | 3 | 2 | 3 | 4 | 5 | 3 | 3 | 0.911413 | 29/30 | NonOverlappingTemplate |
| 4 | 0 | 4 | 6 | 4 | 3 | 0 | 3 | 2 | 4 | 0.299251 | 27/30 | NonOverlappingTemplate |
| 6 | 4 | 1 | 4 | 1 | 2 | 3 | 2 | 3 | 4 | 0.602458 | 25/30 | NonOverlappingTemplate |
| 6 | 3 | 1 | 4 | 3 | 0 | 5 | 3 | 4 | 1 | 0.299251 | 26/30 | NonOverlappingTemplate |
| 5 | 2 | 2 | 4 | 5 | 4 | 2 | 4 | 2 | 0 | 0.534146 | 26/30 | NonOverlappingTemplate |
| 5 | 3 | 1 | 1 | 1 | 5 | 4 | 3 | 5 | 2 | 0.468595 | 29/30 | NonOverlappingTemplate |
| 2 | 2 | 6 | 2 | 7 | 5 | 1 | 3 | 2 | 0 | 0.082177 | 28/30 | NonOverlappingTemplate |
| 1 | 2 | 3 | 3 | 0 | 4 | 7 | 4 | 3 | 3 | 0.299251 | 30/30 | NonOverlappingTemplate |
| 4 | 3 | 1 | 4 | 0 | 4 | 3 | 4 | 3 | 4 | 0.739918 | 27/30 | NonOverlappingTemplate |
| 4 | 1 | 5 | 2 | 3 | 3 | 1 | 3 | 1 | 7 | 0.253551 | 27/30 | NonOverlappingTemplate |
| 2 | 3 | 2 | 3 | 5 | 4 | 3 | 2 | 5 | 1 | 0.804337 | 30/30 | NonOverlappingTemplate |
| 2 | 3 | 1 | 8 | 2 | 3 | 4 | 2 | 1 | 4 | 0.178278 | 30/30 | NonOverlappingTemplate |
| 2 | 1 | 6 | 6 | 1 | 1 | 2 | 5 | 4 | 2 | 0.178278 | 29/30 | NonOverlappingTemplate |
| 4 | 1 | 4 | 3 | 4 | 2 | 4 | 1 | 2 | 5 | 0.739918 | 28/30 | NonOverlappingTemplate |
| 3 | 3 | 1 | 6 | 2 | 1 | 6 | 2 | 3 | 3 | 0.407091 | 30/30 | NonOverlappingTemplate |
| 3 | 5 | 4 | 2 | 5 | 1 | 0 | 4 | 2 | 4 | 0.468595 | 29/30 | NonOverlappingTemplate |
| 1 | 3 | 2 | 4 | 3 | 3 | 3 | 3 | 7 | 1 | 0.468595 | 29/30 | NonOverlappingTemplate |
| 1 | 3 | 5 | 3 | 5 | 1 | 3 | 3 | 2 | 4 | 0.739918 | 30/30 | NonOverlappingTemplate |
| 7 | 2 | 1 | 1 | 2 | 5 | 6 | 2 | 3 | 1 | 0.100508 | 26/30 | NonOverlappingTemplate |
| 3 | 4 | 4 | 1 | 4 | 4 | 4 | 0 | 2 | 4 | 0.671779 | 27/30 | NonOverlappingTemplate |
| 1 | 5 | 3 | 4 | 3 | 3 | 4 | 4 | 1 | 2 | 0.804337 | 30/30 | NonOverlappingTemplate |
| 3 | 4 | 3 | 1 | 3 | 3 | 3 | 2 | 0 | 8 | 0.148094 | 30/30 | NonOverlappingTemplate |
| 1 | 0 | 7 | 6 | 1 | 5 | 7 | 2 | 1 | 0 | 0.002624 | 30/30 | NonOverlappingTemplate |
| 5 | 2 | 3 | 3 | 2 | 4 | 5 | 1 | 2 | 3 | 0.804337 | 27/30 | NonOverlappingTemplate |
| 1 | 0 | 6 | 2 | 3 | 2 | 5 | 6 | 0 | 5 | 0.054199 | 30/30 | NonOverlappingTemplate |
| 1 | 6 | 4 | 0 | 6 | 3 | 4 | 1 | 2 | 3 | 0.178278 | 30/30 | NonOverlappingTemplate |
| 5 | 4 | 4 | 2 | 3 | 0 | 1 | 2 | 7 | 2 | 0.178278 | 28/30 | NonOverlappingTemplate |
| 4 | 4 | 1 | 3 | 4 | 1 | 7 | 2 | 4 | 0 | 0.178278 | 28/30 | NonOverlappingTemplate |
| 6 | 4 | 4 | 1 | 3 | 3 | 3 | 2 | 1 | 3 | 0.671779 | 27/30 | NonOverlappingTemplate |
| 4 | 5 | 2 | 3 | 2 | 6 | 3 | 1 | 1 | 3 | 0.534146 | 28/30 | NonOverlappingTemplate |
| 4 | 3 | 1 | 3 | 0 | 4 | 4 | 4 | 4 | 3 | 0.739918 | 30/30 | NonOverlappingTemplate |
| 1 | 1 | 1 | 8 | 4 | 4 | 2 | 1 | 4 | 4 | 0.082177 | 29/30 | NonOverlappingTemplate |
| 7 | 0 | 1 | 1 | 2 | 4 | 4 | 2 | 3 | 6 | 0.082177 | 27/30 | NonOverlappingTemplate |
| 2 | 2 | 2 | 4 | 2 | 8 | 2 | 4 | 2 | 2 | 0.253551 | 29/30 | NonOverlappingTemplate |
| 4 | 2 | 3 | 4 | 0 | 4 | 3 | 4 | 1 | 5 | 0.602458 | 27/30 | NonOverlappingTemplate |
| 4 | 2 | 4 | 3 | 4 | 4 | 4 | 1 | 3 | 1 | 0.862344 | 28/30 | NonOverlappingTemplate |
| 2 | 3 | 3 | 2 | 2 | 3 | 5 | 2 | 6 | 2 | 0.739918 | 30/30 | NonOverlappingTemplate |
| 1 | 3 | 2 | 9 | 4 | 3 | 1 | 2 | 0 | 5 | 0.017912 | 29/30 | NonOverlappingTemplate |
| 1 | 5 | 3 | 3 | 3 | 0 | 5 | 3 | 4 | 3 | 0.602458 | 30/30 | NonOverlappingTemplate |
| 1 | 0 | 2 | 4 | 2 | 6 | 3 | 2 | 6 | 4 | 0.213309 | 30/30 | NonOverlappingTemplate |
| 3 | 1 | 6 | 4 | 3 | 2 | 0 | 4 | 4 | 3 | 0.468595 | 28/30 | NonOverlappingTemplate |
| 2 | 1 | 4 | 2 | 3 | 4 | 3 | 6 | 3 | 2 | 0.739918 | 29/30 | NonOverlappingTemplate |
| 4 | 3 | 3 | 1 | 3 | 5 | 2 | 2 | 3 | 4 | 0.911413 | 26/30 | NonOverlappingTemplate |
| 3 | 3 | 3 | 2 | 2 | 3 | 5 | 2 | 5 | 2 | 0.911413 | 30/30 | NonOverlappingTemplate |
| 0 | 3 | 3 | 4 | 1 | 5 | 1 | 10 | 3 | 0 | 0.001588 | 30/30 | NonOverlappingTemplate |
| 2 | 5 | 1 | 2 | 1 | 3 | 3 | 5 | 4 | 4 | 0.671779 | 29/30 | NonOverlappingTemplate |
| 2 | 4 | 3 | 4 | 3 | 3 | 4 | 4 | 2 | 1 | 0.949602 | 30/30 | NonOverlappingTemplate |
| 4 | 4 | 4 | 5 | 1 | 6 | 1 | 2 | 2 | 1 | 0.350485 | 28/30 | NonOverlappingTemplate |
| 1 | 5 | 4 | 3 | 3 | 1 | 6 | 5 | 2 | 0 | 0.213309 | 29/30 | NonOverlappingTemplate |
| 2 | 1 | 2 | 5 | 2 | 3 | 2 | 5 | 4 | 4 | 0.739918 | 30/30 | NonOverlappingTemplate |
| 4 | 2 | 0 | 4 | 4 | 0 | 8 | 4 | 0 | 4 | 0.022503 | 27/30 | NonOverlappingTemplate |
| 2 | 6 | 2 | 4 | 1 | 4 | 2 | 2 | 3 | 4 | 0.671779 | 30/30 | NonOverlappingTemplate |
| 1 | 2 | 6 | 2 | 5 | 4 | 3 | 2 | 1 | 4 | 0.468595 | 29/30 | NonOverlappingTemplate |
| 6 | 2 | 2 | 3 | 2 | 1 | 6 | 3 | 3 | 2 | 0.468595 | 26/30 | NonOverlappingTemplate |
| 4 | 2 | 4 | 5 | 2 | 3 | 2 | 4 | 3 | 1 | 0.862344 | 29/30 | NonOverlappingTemplate |
| 2 | 7 | 4 | 3 | 1 | 1 | 2 | 5 | 2 | 3 | 0.299251 | 30/30 | NonOverlappingTemplate |
| 3 | 2 | 4 | 2 | 0 | 5 | 3 | 4 | 6 | 1 | 0.350485 | 27/30 | NonOverlappingTemplate |
| 3 | 2 | 1 | 5 | 3 | 3 | 3 | 6 | 2 | 2 | 0.671779 | 29/30 | NonOverlappingTemplate |
| 1 | 3 | 2 | 3 | 1 | 5 | 8 | 2 | 3 | 2 | 0.148094 | 29/30 | NonOverlappingTemplate |
| 3 | 1 | 1 | 4 | 0 | 7 | 2 | 5 | 3 | 4 | 0.148094 | 30/30 | NonOverlappingTemplate |
| 2 | 3 | 5 | 2 | 1 | 4 | 3 | 2 | 4 | 4 | 0.862344 | 28/30 | OverlappingTemplate |
| 5 | 2 | 5 | 3 | 3 | 3 | 0 | 0 | 1 | 8 | 0.028181 | 28/30 | Universal |
| 2 | 1 | 3 | 2 | 4 | 2 | 5 | 0 | 8 | 3 | 0.082177 | 29/30 | ApproximateEntropy |
| 2 | 2 | 0 | 2 | 1 | 1 | 0 | 1 | 4 | 1 | 0.122325 | 14/14 | RandomExcursions |
| 1 | 4 | 2 | 1 | 1 | 2 | 1 | 1 | 1 | 0 | 0.213309 | 14/14 | RandomExcursions |
| 4 | 0 | 1 | 2 | 3 | 0 | 0 | 1 | 1 | 2 | 0.035174 | 11/14 | RandomExcursions |
| 0 | 1 | 2 | 2 | 1 | 0 | 3 | 2 | 1 | 2 | 0.350485 | 14/14 | RandomExcursions |
| 2 | 1 | 1 | 0 | 3 | 3 | 1 | 1 | 2 | 0 | 0.213309 | 14/14 | RandomExcursions |
| 0 | 1 | 1 | 1 | 3 | 0 | 2 | 2 | 4 | 0 | 0.035174 | 14/14 | RandomExcursions |
| 2 | 1 | 1 | 1 | 2 | 1 | 1 | 1 | 3 | 1 | 0.739918 | 14/14 | RandomExcursions |
| 1 | 1 | 1 | 3 | 2 | 0 | 1 | 1 | 3 | 1 | 0.350485 | 14/14 | RandomExcursions |
| 4 | 1 | 0 | 2 | 1 | 2 | 1 | 1 | 1 | 1 | 0.213309 | 13/14 | RandomExcursionsVariant |
| 2 | 3 | 1 | 0 | 1 | 2 | 3 | 1 | 1 | 0 | 0.213309 | 13/14 | RandomExcursionsVariant |
| 2 | 0 | 2 | 3 | 1 | 2 | 1 | 1 | 0 | 2 | 0.350485 | 12/14 | RandomExcursionsVariant |
| 2 | 0 | 0 | 3 | 3 | 0 | 2 | 1 | 2 | 1 | 0.122325 | 13/14 | RandomExcursionsVariant |
| 2 | 0 | 2 | 1 | 1 | 0 | 4 | 1 | 2 | 1 | 0.122325 | 12/14 | RandomExcursionsVariant |
| 4 | 1 | 1 | 0 | 0 | 1 | 1 | 3 | 2 | 1 | 0.066882 | 12/14 | RandomExcursionsVariant |
| 4 | 1 | 1 | 1 | 1 | 0 | 1 | 4 | 1 | 0 | 0.017912 | 12/14 | RandomExcursionsVariant |
| 3 | 1 | 3 | 1 | 1 | 2 | 0 | 1 | 1 | 1 | 0.350485 | 12/14 | RandomExcursionsVariant |
| 1 | 3 | 3 | 1 | 1 | 0 | 2 | 1 | 1 | 1 | 0.350485 | 14/14 | RandomExcursionsVariant |
| 1 | 1 | 2 | 1 | 1 | 1 | 1 | 2 | 4 | 0 | 0.213309 | 14/14 | RandomExcursionsVariant |
| 2 | 0 | 2 | 3 | 1 | 0 | 1 | 2 | 2 | 1 | 0.350485 | 14/14 | RandomExcursionsVariant |
| 1 | 2 | 1 | 1 | 3 | 0 | 3 | 1 | 1 | 1 | 0.350485 | 14/14 | RandomExcursionsVariant |
| 1 | 1 | 1 | 3 | 2 | 1 | 2 | 1 | 0 | 2 | 0.534146 | 13/14 | RandomExcursionsVariant |
| 0 | 2 | 0 | 1 | 4 | 1 | 3 | 0 | 3 | 0 | 0.008879 | 14/14 | RandomExcursionsVariant |
| 1 | 0 | 2 | 2 | 0 | 2 | 1 | 1 | 1 | 4 | 0.122325 | 14/14 | RandomExcursionsVariant |
| 1 | 2 | 0 | 1 | 1 | 0 | 2 | 1 | 4 | 2 | 0.122325 | 13/14 | RandomExcursionsVariant |
| 1 | 1 | 2 | 0 | 2 | 0 | 2 | 3 | 1 | 2 | 0.350485 | 13/14 | RandomExcursionsVariant |
| 1 | 0 | 2 | 1 | 3 | 3 | 0 | 2 | 1 | 1 | 0.213309 | 14/14 | RandomExcursionsVariant |
| 7 | 2 | 1 | 4 | 4 | 2 | 5 | 2 | 2 | 1 | 0.253551 | 26/30 | Serial |
| 2 | 2 | 4 | 3 | 1 | 4 | 5 | 4 | 2 | 3 | 0.862344 | 29/30 | Serial |
| 3 | 2 | 2 | 3 | 3 | 5 | 2 | 5 | 2 | 3 | 0.911413 | 29/30 | LinearComplexity |

The minimum pass rate for each statistical test with the exception of the exception of the random excursion (variant) test is approximately = 24 for a sample size = 30 binary sequences.

The minimum pass rate for the random excursion (variant) test is approximately 10 for a sample size = 14 binary sequences.